\documentclass[12pt]{amsart}
\usepackage{amscd}
\begin{document}
\begin{titlepage}
\begin{center}

{\large \bf Some Global Aspects of Gauge Anomalies of Semisimple Gauge Groups
and Fermion Generations in GUT and Superstring Theories}
\\[.1in]
H. ZHANG \\[.1in]
{\em 207 Admiral Lane, Greer, SC 29650 U.S.A.}
\footnote{e-mail:john.johnz@gmail.com, 
Postal address:104 Tudor Lane, Middle Island, NY 11953 U.S.A.}



{\bf Abstract}
\end{center}
We study more extensively and completely for global gauge anomalies
with some semisimple gauge groups as initiated in ref.1.
A detailed and complete proof or derivation is provided for
the $Z_2$ global (non-perturbative)
gauge anomaly given in ref.1 for a gauge theory with
the semisimple gauge group $SU(2)\times SU(2)\times SU(2)$ in
$D=4$ dimensions and Weyl fermions in the irreducible representation (IR)
$\omega=(2,2,2)$ with 2 denoting the corresponding dimensions.
This $Z_2$ anomaly was
used in the discussions related to generic $SO(10)$ and supersymmetric
$SO(10)$ unification theories$^1$ for the total generation numbers of fermions
and mirror fermions.  Our result$^1$ that the global anomaly coefficient
formula is given by $A(\omega)=exp[i{\pi}Q_2(\Box)]=-1$ in this case with
$Q_2(\Box)$ being the Dynkin index for $SU(8)$ in the fundamental IR
$(\Box)=(8)$ and that the corresponding gauge
transformations need to be topologically non-trivial simultaneously in all the
three $SU(2)$ factors for the homotopy group
$\Pi_4(SU(2)\times SU(2)\times SU(2))$ is also discussed, and
as shown by our results$^1$ the semisimple gauge transformations collectively
may have physical consequences which do not correspond to successive
simple gauge transformations.
The similar result given in ref.1
for the $Z_2$ global gauge anomaly of gauge group $SU(2)\times SU(2)$
with Weyl fermions in the IR $\omega=(2,2)$ with 2 denoting the corresponding
dimensions is also discussed with proof similar to the case of
$SU(2)\times SU(2)\times SU(2)$.
We also give a complete proof for some relevent topological results.
We expect that our results and discussions may
also be useful in more general studies related to global aspects of
gauge theories. Gauge anomalies for the relevant semisimple gauge groups
are also briefly discussed in higher dimensions, especially for
self-contragredient representations, with discussions involving
trace identities relating to ref.14.
We also relate the discussions to our results and propositions
in our previous studies of global gauge anomalies.
We also remark the connection of our
results and discussions to the total generation numbers in relevant theories.

\end{titlepage}
\newpage
\setcounter{page}{1}
\section{\bf Introduction}
Gauge symmetries have been a crucial aspect in our progress of 
understanding the fundamental interactions.  
Analysis of symmetries and intrinsic consistency may often provide useful 
and important hints as well as constraints. An outstanding example
is the theory of anomalies due to its remarkable role in  
model buildings of fundamental interactions and their unifications. 
History seems indicating that new knowledges and understandings in 
this area may be of fundamental importance. 

A main purpose of this paper is to provide a complete proof or
derivation for the $Z_2$ global (non-perturbative) 
gauge anomalies$^1$ for gauge theories in D=4 dimensions with the 
semisimple gauge group 
$SU(2)\times SU(2)\times SU(2)$ (or $SU(2)\times SU(2)$) 
with Weyl fermions in the irreducible representation (IR)
$\omega=(2,2,2)$ (or $\omega=(2,2)$), where each 2 in the $\omega$ will denote the corresponding 
dimensions for each $SU(2)$ hereafter. This $Z_2$ global anomaly is
stated as the Proposition 5 in ref.1.
In our discussions, the $SU(2)\times SU(2)\times SU(2)$ will also be 
denoted by ${(SU(2))}^3$. As it can be seen from
our results and detailed analysis that the simple factors in our semisimple 
gauge groups can act collectively to generate important physical consequences
which do not correspond to successive simple gauge transformations.  
We also have the $Z_2$ global gauge anomaly similarly for
the semisimple gauge group 
$SU(2)\times SU(2)$ in D=4 dimensions with Weyl fermions in the IR 
$\omega=(2,2)$, and this $Z_2$ global anomlay is stated as
the Proposition 6 in ref.1.
For our purpose, we will focuse on the global (non-perturbative)
gauge anomalies for the semisimple gauge groups, especially the ${(SU(2))}^3$. 
The result and analysis for $SU(2)\times SU(2)$ 
are paralell to that for the ${(SU(2))}^3$ for which 
we will present in more details. 
Our method will be topological. Especially, we will use 
diagram-chasing method in algebraic topology. 
With this method, some relevant and useful toplogical results are proved 
completely, and then the topological results will be used to provide a
rigorous and complete proof for the $Z_2$ anomalies. 
We will also relate our discussions to our results and propositions
with our previous studies of global gauge anomalies.
 
Since it was noted by Witten$^2$ that an $SU(2)$ gauge theory in four 
dimensions with an odd number of left chiral doublets of fermions 
is mathematically inconsistent, global (non-perturbative) gauge anomalies 
have been studied rather systematically and extensively. 
It is also known$^{3,4}$ that global anomalies in $D=2n$ 
dimensions, including gravitation may be generally expressed in terms of 
Atiyah-Singer index for the Dirac operator in 2n-dimensional space M and an
integral including the Dirac genus ${\hat A}$ and the Chern character ChF in 
a (2n+2)-dimensional space. However, for pure gauge anomalies, it
is often more convenient to compute the coefficient $A(\omega)$ of pure global
gauge anomaly for a representation (rep) $\omega$ by another method due to
Elizur and Nair$^5$. This method was utilized by many authors$^6$ for
 some cases. It has been also utilized with additional
topological and Lie-algebraic methods in refs.7-12 to determine 
for the possibilities of global gauge anomalies in very general cases.  
Especially, many general and systematic results for 
$A(\omega)$ in 2n dimensions  
were obtained in refs. 7 and 8 for $SU(N)$ groups and other simple groups
in $D=2n$ dimensions. More studies are also given in refs.9-12,1 in $D=2n$ 
dimensions. For instance, as $SU(2)$ and complex Stiefel manifolds
are also intimately connected to our discussions here, 
we note that the possible $SU(2)$ global gauge anomalies
were determined completely in arbitrary $2n$ dimensions, and $SU(N)$ global 
gauge anomlies for general N were studied rigorously with 
the global anomaly coefficient $A(\omega)$ for a rep $\omega$ of
$SU(n-k)$ $(0\le k\le n-2)$ in $D=2n$ expressed as    
the general formula
\begin{equation}
A(\omega)=exp\{\frac{2\pi i}{d_{n+1,k+1}}Q_{n+1}(\tilde {\omega})\}.     
\end{equation}
Where in the above formula, the
\begin{equation}
d_{n+1,k+1}=\frac{n!}{U(n+1,k+1)}=integers,                      
\end{equation}
is the James number$^{13}$ for the complex Stiefel manifold
$SU(n+1)/SU(n-k)$, and the $Q_{n+1}(\tilde {\omega})$ is the (n+1)-th 
generalized Dynkin index$^{14}$ for the rep $\tilde {\omega}$ of the $SU(n+1)$
group. The rep $\tilde {\omega}$ of $SU(n+1)$ must satisfy the requirement that
under the reduction of $SU(n+1)$ into $SU(n-k)$ the rep $\tilde {\omega}$ 
reduces to a direct sum of $\omega$ and $SU(n-k)$ singlets allowing possible
negative multiplets for opposite chirality.

Global gauge anomalies for semisimle gauge groups were 
initially studied in ref.1 where some results related to the possibilities of
global gauge anomalies for semisimple gauge groups were given and studied. 
In this paper, we will further and rigorously 
study some topological aspects related to some of the results and propositions. 

Our approach will be using exact homotopy sequences of fiber bundles and
diagram-chasing related to commutative diagrams for homomorphisims between
the exact homotopy sequences.   
We will also utilize some of the discussions and results in refs.7-8 etc. 
for our consideration and analysis of the global gauge anomalies here for
the semisimple gauge groups ${(SU(2))}^3$ etc. As we will see that a 
very useful topological result needed for our purpose of analysizing the group
${(SU(2))}^3$ in four dimensions is for the homotopy group
\begin{equation}
\Pi_{5}(SU(8)/{(SU(2))}^3) = Z\oplus Z_2\oplus Z_2 ,
\label{3} 
\end{equation}
which was first given in ref.1 without proof. We will present a proof of
the above result in this paper. 
We also have the similar result in the other case
\begin{equation}
\Pi_{5}(SU(4)/{(SU(2))}^2) = Z\oplus Z_2 , 
\label{4} 
\end{equation}
used in ref.1 for ${(SU(2))}^2$.
We will focus our analysis on the semisimple gauge group ${(SU(2))}^3$,  
as the analysis for the ${(SU(2))}^2$ can be rather similar or parallel
as we will see. 
 
First of all, since the theory will not be consistent if
a gauge group $H$ under consideration for global gauge anomaly has a local 
anomaly, we assume that the rep $\omega$ of $H$ in our discussions with
strong anomaly-free condition instead of Green-Schwarz mechanism obey
$Tr^{(\omega)}F^{(n+1)}=0$. In particular we must have 
\begin{equation}
Tr^{(\omega)}F^{3} = 0 ,  
\end{equation}
with $n=2$ for our pure gauge theory in four dimensions. 
Here F denotes the field-strength differential two-forms of values in 
Lie algebra of $H$. As it is emphasized that$^7$, with the strong anomaly-free 
condition, the global gauge anomaly coefficient is given by 
$A(\omega)=(-1)^{ind D_{2n}}$, where the $ind D_{2n}$ stands for the 
Atiya-Singer index of Dirac operator in $D=2n$ dimensions, so that the global
gauge anomaly is at most of like a $Z_2$ type.

In our discussions, note that for the Lie groups $H$ and $G$ with $H\subset G$,
 the G can be considered as principal bundle over base space $G/H$ with fiber 
$H$. Therefore, we can consider the following exact homotopy sequence$^{15}$

\begin{equation}
\Pi_{2n+1}(G)\stackrel{P_*}{\longrightarrow}\Pi_{2n+1}(G/H)
\stackrel{\Delta_*}{\longrightarrow}\Pi_{2n}(H)\stackrel{i_*}
{\longrightarrow}\Pi_{2n}(G) = 0 .
\end{equation}

In $D=2n=4$ dimensions, it can be written as 
\begin{equation}
\Pi_{5}(G)\stackrel{P_*}{\longrightarrow}\Pi_{5}(G/H)
\stackrel{\Delta_*}{\longrightarrow}
\Pi_{4}(H)\stackrel{i_*}{\longrightarrow}\Pi_{4}(G) = 0 .
\end{equation}

For the relevant gauge groups $H$ and $G$ we focus in this paper, as it will be
seen that we will have
$\Pi_{2n+1}(G)=Z$ and $\Pi_{2n+1}(G/H)=Z\oplus T$, where $T$ is a torsion group.
 
We will organize our paper as follows. In the next section, we will use the 
as an assumption the topological result ${P_*}(x) = 2y + t$ for the specific 
gauge groups $H={(SU(2))}^3$ or ($H={(SU(2))}^2$) and $G=SU(8)$ 
(or $G=SU(4)$) in this paper, 
where $x\in \Pi_{2n+1}(G)$ and $y\in \Pi_{2n+1}(G/H)$ are for generating 
elements of the corresponding $Z$'s and $t$ is a possible torsion element 
in $\Pi_{2n+1}(G/H)$, i.e. we assume the homotopy group for 
$\Pi_{5}(SU(8)/{(SU(2))}^3)$ in Eq.(3) and similarly with Eq.(4). 
We will see the relevant $Z_2$ global gauge anomalies for the semisimple gauge 
groups using our previous results given as propositions. 
After section 2, we will then be fully motivated for the section 3 to
provide a complete proof of the topological results we assumed in section 2. 
Therefore, the $Z_2$ global anomalies in our focus are then rigorously and
completely
proved when the topologcal results used to show the $Z_2$ anomalies are proved. 
Some useful topological results more general than what we assumed will be 
proved in the process, and we expect that these results may also be useful 
for more general or further study of global gauge anomalies. 
We will also discuss briefly in section 4 about global gauge anomalies for 
some semisimple gauge groups in higher dimensions, with discussions involving
trace identities relating to ref.14. We section 4, we also give a proposition 
for the absence of global gauge anomalies for semisimple gauge groups 
in self-contragredient representations 
in arbitrary $D=4k+2$ dimensions satisfying strong anomaly-free conditions
of local (perturbative) gauge anomalies.  
We also remark the connection of our results and discussions to 
the total generation numbers in the relevant theories such as GUST or
Superstring theories.
We will also provide an appendix with some relevant and useful theorems etc. 
for the convenience of our discussions as well as topological proofs.

\section{\bf The $Z_2$ Global Gauge Anomalies for the Semisimple Gauge Groups and Our Previous Results} 

In our consideration of the gauge group 
$H={(SU(2))}^3$ in the rep 
$\omega=(2,2,2)$ with each 2 denoting the corresponding dimensions.
Note that the gauge group $H$ can be embedded into the simple 
gauge group $G=SU(8)$ in the fundamental rep $\tilde{\omega}=(\Box)$ or (8) 
with 8 denoting the dimensions. We now have that $\Pi_4(G)=0$ being trivial and
also that the $\tilde{\omega}$ reduces to ${\omega}$ as the $G=SU(8)$ reduces to
the $H={(SU(2))}^3$. Furthermore, the local anomaly-free
condition Eq.(5) is automatically obeyed for the $H$.  
Therefore, the possible global gauge
anomaly for the gauge group $H={(SU(2))}^3$ may be studied
as the local gauge anomalies in the $G=SU(8)$. 

Now for $H={(SU(2))}^3$, we have
\begin{equation}
\Pi_4({(SU(2))}^3) = Z_2\oplus Z_2\oplus Z_2 .
\end{equation}  
Therefore, for our consideration, we can write Eq.(7) as
\begin{equation}
\Pi_{5}(SU(8))\stackrel{P_*}{\rightarrow}
\Pi_{5}(SU(8)/{(SU(2))}^3)
\stackrel{\Delta_*}{\rightarrow}\Pi_{4}({(SU(2))}^3)
\stackrel{i_*}{\rightarrow}\Pi_{4}(SU(8)) = 0 .
\end{equation} 
We then have$^1$
\begin{equation}
Z\stackrel{P_*}{\longrightarrow}
Z\oplus Z_2\oplus Z_2
\stackrel{\Delta_*}{\longrightarrow}Z_2\oplus Z_2\oplus Z_2
\stackrel{i_*}{\longrightarrow} 0 , 
\end{equation}
with$^{15}$ $\Pi_{5}(SU(8))=Z$ and$^{1}$
\begin{equation}
\Pi_{5}(SU(8)/{(SU(2))}^3) = Z\oplus Z_2\oplus Z_2 .
\end{equation}

Similarly for $H={(SU(2))}^2$ and $G=SU(4)$, we have correspondingly
\begin{equation}
Z\stackrel{P_*}{\longrightarrow}
Z\oplus Z_2
\stackrel{\Delta_*}{\longrightarrow}Z_2\oplus Z_2
\stackrel{i_*}{\longrightarrow} 0 , 
\end{equation}
 
To proceed, the Proposition 2 in ref.8 will be useful. It is stated as 
follows:

Suppose that we have $\Pi_{2n+1}(G)=Z$ and $\Pi_{2n+1}(G/H)=Z\oplus T$,
where $T$ is a (finite) torsion group. Let $x\in \Pi_{2n+1}(G)$ and
$y\in \Pi_{2n+1}(G/H)$ be generating elements of the $Z$'s. Suppose that
they are related by
\begin{equation}
{P_*}(x) = dy + t ,
\end{equation}
where d is a nonzero integer and t is an element of the $T$.
Then, the global anomaly coefficient $A(\omega)$ of $H$ is effectively given
by
\begin{equation}
A(\omega) = exp[{\frac{2\pi i}{d}}Q_{n+1}(\tilde {\omega})] ,
\end{equation}
in a sense that all other global anomalies are some integral powers of the 
express $A(\omega)$ given above. This formula may need to be slightly
modified for $G=SO(2n+2)$ with $n=odd$ for reasons explained in ref.8, but
it will not so relevent to our discussions here. 
We note here that Eq.(1) is similar to the form of the Eq.(14).

In the particular cases we are interested in here,
we will provide a detailed toplogical proof in the next section 
that we have $d=2$ leading to the
 $Z_2$ global gauge anomalies in $D=4$ dimensions for the relevant semisimple 
gauge groups, together with Eq.(3) for homotopy group 
$\Pi_{5}(SU(8)/{(SU(2))}^3) = Z\oplus Z_2\oplus Z_2$. In this section,
we will use the two results respectively as an assumption. 
 
We also like to note that a useful result given in Proposition 4 in refs.9-10 
or in ref.8 (section II) is consistent with
the two results $d=2$ in this case for Eq.(14)
and the homotopy group Eq.(3) as an assumption here. 
That result gives as
\begin{equation}
d = {\frac{ord[\Pi_{2n}(H)]}{ord(T)}},
\end{equation}
with $ord(T)$ denoting the order of a finite group $T$ and $ord(T)=1$ when T is
empty. Note here that $\Pi_{2n}(H)$ is always finite by a theorem proved by
Serre$^{16}$. 
Hence, the anomaly coefficient $A(\omega)$ with the above Eq. may also be 
written as 
\begin{equation}
A(\omega) = exp\{{2\pi i}{\frac{ord(T)}{ord[\Pi_{2n}(H)]}}Q_{n+1}
(\tilde {\omega})\} . 
\end{equation}
The consistency can be easily seen as follows.
For Eq.(9) or Eq.(10) in our consideration in $D=4$ dimensions, 
with $T=Z_2\oplus Z_2$,
$\Pi_4(H)=Z_2\oplus Z_2\oplus Z_2$ and $\tilde{\omega}=(\Box)$, it is then 
obvious that Eq.(15) gives $d=2$. 

With our assumption of $d=2$, the anomaly
coefficient formula Eq.(14) is then given by
\begin{equation}
A(\omega) = exp[i{\pi}Q_3(\Box)], 
\end{equation}
or
\begin{equation}
A(\omega) = exp[i{\pi}Q_2(\Box)] = -1,
\end{equation}
as given in ref.1 with $Q_2(\Box)=1$ and the even-odd rule$^{10,17,18}$. 
Where the even-odd rule is written generally as$^{10}$ 
\begin{equation}
Q_k(\omega) = Q_2(\omega) (mod 2),
\end{equation}
for a rep $\omega$ of $SU(N) (2\leq k\leq N)$ or $SP(2N) (2\leq k\leq 2N)$ or
$SO(2N+1)$ or $SO(2N) (2\leq k\leq 2(N-1) (k=even))$. The general form of the
even-odd rule in the case of $SU(N) (N\geq 3)$ for $k=3$ above reduces to the
special case of refs.17-18.   
Therefore, with $Q_2(\Box)=1$ for $SU(8)$ and the assumption of the topological
result which will be proven in details in the next section, 
the anomaly coefficient 
in our consideration of $H={(SU(2))}^3$ in the rep $\omega=(2,2,2)$ 
with 2 denoting the corresponding dimensions is $A(\omega)=-1$, so that 
it has a $Z_2$ global gauge anomaly. 

Having seen the $Z_2$ global gauge anomaly in the above, we can also see 
the corresponding element in the homotopy group 
$\Pi_4(H)=Z_2\oplus Z_2\oplus Z_2$
for the anomaly $A(\omega)=-1$.
Using Eq.(13) with $d=2$, we obtain
\begin{equation}
{P_*}(x) = 2y + t . 
\end{equation}
Where $x\in \Pi_{5}(G)=Z$ and $y\in \Pi_{5}(G/H)=Z\oplus T$ are the 
generating elements of
the $Z's$ with $G=SU(8)$ and $H={(SU(2))}^3$,  
$t\in T=Z_2\oplus Z_2$ is a torsion element. 
The image of the $P_*$ is then given by
\begin{equation}
Im{P_*} = \bigcup \{(2ky + kt), k=integers\in Z\} . 
\end{equation}
Since $Im{P_*}=Ker{\Delta_*}$ by the exactness of the homotopy sequence$^{15}$
\begin{equation}
Ker{\Delta_*} = \bigcup \{(2ky + kt) , k=integers\in Z\} ,
\end{equation}
i.e. only the elements of the form $2ky+kt$ for integers k can be mapped to 
the identity element of $\Pi_4(H)=Z_2\oplus Z_2\oplus Z_2$. This obviously
implies that the elements
\begin{equation}
{\Delta_*}[(2k+1)y + t^{\prime}] \neq 0 , 
\end{equation}
for any $t^{\prime}\in T$ and integer k. In particular, we have
\begin{equation}
{\Delta_*}[y + t^{\prime}] \neq 0 ,
\end{equation}
for any $t^{\prime}{\in T}$.  
Therefore$^1$, the generating element of $Z$ in
$\Pi_{5}(SU(8)/{(SU(2))}^3)$
must be mapped to a non-trivial element in the
$\Pi_{4}({(SU(2))}^3)$,
which is $Z_2\oplus Z_2\oplus Z_2$ in the canonical form as an abelian group
and the three ${Z_2}$'s are symmetric as we emphasized$^1$. 
This canonical form is unique up to a rearrangement of the $Z_2$'s
(see Appendix for canonical form). 
Therefore, up to at most a possible torsion element, we obtain evidently
\begin{equation}
{\Delta_*}(y) = \{1,1,1\} , 
\end{equation}
with the 1 being the non-trivial element of the $Z_2$.
The above equation is our assertion given in ref.1 that the corresponding 
anomalous gauge transformations need to be topologically non-trivial 
simultaneously in all the three $SU(2)$ factors for $\Pi_{4}({(SU(2))}^3)$. 
Before going to the next section for more detailed topological proof, 
we will give some relevant remarks.

{\bf Remark (1)}: The global gauge anomaly here for the 
${(SU(2))}^3$ is in a sense different from the global
gauge anomaly for a simple gauge group as the $SU(2)$ global gauge anomaly
noted first. In our case, if only one or two of the $SU(2)$ factors are
topologically non-trivial for it's forth homotopy group, it is not sufficient 
for the anomaly to appear. Instead, only when a gauge transformation is 
topologically non-trivial simultaneously in all the three $SU(2)$ factors
for it's forth homotopy group, the global gauge anomaly is obtained.
As emphasized in ref.1 that in our case when it is topologically non-trivial 
simultaneously in all the three $SU(2)$ factors for the forth homotopy group, 
the 
anomaly coefficient will not get any trace factor 2 in the phase from any of 
the $SU(2)$'s. 
Our results showed that the simple ideals of a semisimple gauge groups
may act together like a whole entity or symmetrically to generate non-trivial 
physical consequences which may not be obtained by the individual or sequential
actions of the simple ideals. 

{\bf Remark (2)}: As given in ref.1 that $SU(2)\times SU(2)$ in the rep
$\omega=(2,2)$ with 2 denoting the corresponding dimensions also has a $Z_2$
global gauge anomaly in four dimensions, this can be also seen similarly using 
Eq.(4)
with $G=SU(4)$ in $\tilde{\omega}=(\Box)$ or (4) for dimensions. Again,
the $SU(2)$'s need to be topologically non-trivial simultaneously in
both $SU(2)$ factors for the forth homotopy group to generate the
global anomaly. In this section, we can see the $Z_2$ anomaly with
the assumption of $d=2$ for Eq.(13) or Eq.(14), and we also note that
Eq.(4) and Eq.(15) are consistent with this.

{\bf Remark (3)}: As we emphasized in ref.1 that the both cases of
$H=SU(2)\times SU(2)$ and $H={(SU(2))}^3$ for the
$Z_2$ anomaly is also due to the fact that the $\Pi_4(H)$ has one more
$Z_2$ than the torsion of $\Pi_5(G/H)$.
 
As we have seen that the result $d=2$ or the relevant homotopy group $\Pi_4(H)$ 
 is crutial for the $Z_2$ global anomaly. 
As indicated, we are now fully motivated for the next section to provide a
complete topological
proof of $d=2$ and the relevant homotopty group $\Pi_4(H)$ 
such as Eq.(3) for the $\Pi_{5}(SU(8)/{(SU(2))}^3)$.

\section{\bf Proof of the Topological Results and Homotopy Groups}
In this section, we will prove the toplogical result stated in the section 1 and
also expressed in Eq.(20) which is Eq.(13) with $d=2$ and
the homotopy group given by Eq.(3) with $H={(SU(2))}^3$ and $G=SU(8)$. 
The proof of $d=2$ is similar with $H={(SU(2))}^2$, $G=SU(4)$ and Eq.(4).
For the sake of notations, the $H$ and $G$ will also be used for other
groups with clarifications. 

We will use the method of diagram-chasing in algebraic topology.
In our considerations, we will especially use commutative diagrams
related to homomorphisms between exact homotopy sequences of fiber bundles.

As a preparation, we first note a known fact$^{15}$ as follows.
Let $\beta=\{B,P,X,Y,H\}$ be a bundle with bundle space B, base space X, fiber Y, group H, and projection P, let Y be the fiber over $x_0\in X$ and $y_0\in Y$.
Let $\beta^{\prime}={B^{\prime},P^{\prime},X^{\prime},Y^{\prime},H^{\prime}}$
be another bundle, and $h$ be a map of $\beta$ into $\beta^{\prime}$.
Let $\bar{h}: X\longrightarrow X^{\prime}$ with $\bar{h}(x_0)={x}^{\prime}_0$ 
be the induced map of base space, let $h_0: Y_0\longrightarrow Y^{\prime}_0$
be $h\arrowvert Y_0$ for the fibers at $x_0$ and ${x}^{\prime}_0$ respectively.
We obtain then a homomorphism of the homotopy sequence of $\beta$ at $y_0$ 
into that of $\beta^{\prime}$ at ${y_0}^{\prime}=h(y_0)$ 

$$
\CD
\Pi_{N}(B) @>P_*>>\Pi_{N}(X) @>{\Delta_*}>>\Pi_{N-1}(Y_0) @>i_*>>\Pi_{N-1}(B
) \\
@V h_* VV @V \bar{h}_* VV @V {h_0}_* VV @V h_* VV\\
\Pi_{N}(B^\prime) @>{P^\prime}_*>>\Pi_{N}(X^\prime) @>{{\Delta^\prime}_*}>>
\Pi_{N-1}({Y^\prime}_0) @>{i^\prime}_*>>\Pi_{N-1}(B) \\
\endCD
$$

and this diagram is commutative$^{15}$. We note that the exact sequences in 
each row can be in longer form although only the short form was written above. 

For our purpose, we are interested in such commutative diagrams for principal 
bundles with Lie groups for which the relevant homotopy groups based upon
different points are isomorphic.
We notice here that we have used a class of  
commutative diagrams for the homomorphisms
between two homotopy seqences with $G\supset{H^\prime}\supset{H}$
given as

$$
\CD
\Pi_{2n+1}(G) @>P_*>>\Pi_{2n+1}(G/H) @>{\Delta_*}>>\Pi_{2n}(H) @>i_*>>\Pi_{2n}(G) \\
@V i VV @V q_* VV @V {q^\prime}_* VV @V i VV\\
\Pi_{2n+1}(G) @>{P^\prime}_*>>\Pi_{2n+1}(G/{H^\prime}) @>{{\Delta^\prime}_*}>>\Pi_{2n}(H^\prime) @>{i^\prime}_*>>\Pi_{2n}(G) \\
\endCD
$$

especially in the case of $G=SU(n+1), H^\prime=SU(n)$, and 
$H=SU(n-k) (n-k\geq 2)$ with the quotient group $G/H$ 
as the complex Stiefel manifold $SU(n+1)/SU(n-k)$ which we denote 
by ($W_{n+1,k+1}$ (in ref.8) or $O_{n+1,k+1}(C)$ (in ref.9)) 
$V_{k+1}(C^{n+1})$ here for convenience. 
In this case, we actually have both $B$ and $B^\prime$ as $G$ 
or $SU(n+1)$ for the study of simple gauge groups.  
In our consideration of semisimple gauge groups, we will involve the principal
bundles with $G$ and $G^\prime$ being different as we will see. 

As it was indicated, for more general gauge groups $H\subset G$,
we have Eq.(6) for the fibration corresponding to the
principal bundle $G$ over base space $G/H$ and fiber $H$.
In the proceeding proof, we need the relevant exact homotopy sequence
in a longer form

\begin{equation}
\Pi_{2n+1}(H)\stackrel{i_*}{\longrightarrow}
\Pi_{2n+1}(G)\stackrel{P_*}{\longrightarrow}\Pi_{2n+1}(G/H)
\stackrel{\Delta_*}{\longrightarrow}\Pi_{2n}(H)\stackrel{i_*}
{\longrightarrow}\Pi_{2n}(G) = 0 .
\end{equation}

We will also use more specific notations of the mappings for 
different specific cases, and we may also omit some notations for
simplicity.

Firt of all, for $H=SU(2), G=SU(3)$ in $D=2n=4$ dimensions, we have

\begin{equation}
\Pi_{5}(SU(2))\stackrel{{i}_*}{\longrightarrow}
\Pi_{5}(SU(3))\stackrel{{P_3}_*}{\longrightarrow}\Pi_{5}(SU(3)/SU(2))
\stackrel{\Delta_*}{\longrightarrow}\Pi_{4}(SU(2))
{\longrightarrow}\Pi_{4}(SU(3)) = 0 . 
\end{equation}

With $SU(3)/SU(2)=S^5$ and the relevant homotopy groups, 
this gives

\begin{equation}
Z_2 \stackrel{{i}_*}{\longrightarrow} Z\stackrel{{P_3}_*}{\longrightarrow} Z
{\longrightarrow} Z_2 {\longrightarrow} 0 .
\end{equation}

Since $Z_2$ is a torsion group, for the above particular mapping ${i}_*$, there is no non-trivial homomorphism, so that ${i}_* = 0$, we then have

\begin{equation}
0 {\longrightarrow} Z\stackrel{{P_3}_*}{\longrightarrow} Z
{\longrightarrow} Z_2 {\longrightarrow} 0 .
\end{equation}

It can be easily seen by the exactness of the above homotopy sequence that

\begin{equation}
{P_3}_*(x) = 2y ,
\end{equation}

with $x$ and $y$ being the generators of the $\Pi_{5}(SU(3))=Z$ and 
$\Pi_{5}(SU(3)/SU(2))=Z$ respectively. 

With the above result, and $\Pi_{5}(SU(N))=Z for (N\geq 3)$, We will next 
prove a more general result which may be stated as the following theorem.

{\bf Theorem 1}. {\it Consider the Complex Stiefel manifold 
$V_{N-2}(C^N)=SU(N)/SU(2) (N\geq 3)$,
and the exact homotopy sequence
\begin{equation}
\Pi_{5}(SU(2))\stackrel{i_*}{\longrightarrow}
\Pi_{5}(SU(N))\stackrel{{P_N}_*}{\longrightarrow}\Pi_{5}(V_{N-2}(C^N))
\stackrel{\Delta_*}{\longrightarrow}\Pi_{4}(SU(2))
{\longrightarrow}\Pi_{4}(SU(N)) = 0 , 
\end{equation}

or written as

\begin{equation}
0{\longrightarrow} Z\stackrel{{P_N}_*}{\longrightarrow}\Pi_{5}(V_{N-2}(C^N))
\stackrel{\Delta_*}{\longrightarrow} Z_2
{\longrightarrow} 0 ,
\end{equation}

with ${i}_* = 0$ effectively and a finite $\Pi_{5}(SU(2))$. 
Let $x$ be a generator of $\Pi_{5}(SU(N))=Z$. Then, we have 
\begin{equation}
{P_N}_*(x) = 2y ,
\end{equation}
up to a possible torsion element of $\Pi_{5}(SU(N)/SU(2))=Z\oplus T$ with
y denoting the generator of the $Z$. 
 }  

We can provide the proof as follows. 
We have seen the result for for $V_1(C^3)=SU(3)/SU(2)$ with $N=3$.
In the case of $V_2(C^4)=SU(4)/SU(2)$ with $N=4$, we have the 
commutative diagram given by

$$
\CD
0 @>>> \Pi_{5}(SU(3)) @> {P_3}_*>> \Pi_5(V_1(C^3)) @>>> \Pi_{4}(SU(2)) @>>> 0 \\
@. @V {i_3}_* VV @V {q_3}_* VV @V {i_d}_* VV\\
0 @>>> \Pi_{5}(SU(4)) @> {P_4}_*>> \Pi_5(V_2(C^4)) @>>> \Pi_{4}(SU(2)) @>>> 0 \\
\endCD
$$

or

$$
\CD
0 @>>> Z @> {P_3}_*>> Z @>>> Z_2 @>>> 0 \\
@. @V {i_3}_* VV @V {q_3}_* VV @V {i_d}_* VV\\
0 @>>> Z @> {P_4}_*>> \Pi_5(V_2(C^4)) @>>> Z_2 @>>> 0 \\
\endCD
$$

where ${i_N}_*$ denotes the induced isomorphism by the inclusion map 
$i_N: SU(N) \rightarrow SU(N+1)$, and the identification map induced
${i_d}_*$ is also an isomorphism. By Five lemma or Steenrod five lemma
(see Appendix), the ${q_3}_*$ is also an isomorphism. 
With ${P_3}_*(x)=2y$ or Eq.(31), and
the communitativity of the diagram, we can also write

\begin{equation}
{P_4}_*(x) = 2y ,
\end{equation}

up to a possible torsion element with $x$ and $y$ being generators of 
the $\Pi_{5}(SU(4))=Z$ and the $Z$ in $\Pi_5(SU(4)/SU(2))$ respectively.
One can actually also see the known result 
$\Pi_5(V_2(C^4))=\Pi_5(SU(4)/SU(2))=Z\oplus T$
for a torsion $T$ with tensor product of the commutative diagram by the
additive group $Q$ of the rational numbers. 
We have then proved the above theorem 1. We will give more details
for such tensor product later related to the commutative diagram 
involving $SU(8)/{(SU(2)}^3)$ etc.   

For the larger $N$'s, we can see the result similarly. Obviously, we can 
also use the compositions of ${i_N}_*$
or by induction to see ${P_N}_*(x) = 2x$ for Eq.(33) in the theorem.
We also see the result

\begin{equation}
\Pi_5(SU(N)/SU(2))=Z\oplus T,
\end{equation}
for a torsion $T$ which is dependent on N. Such relevant $Z\oplus T$ homotopy
groups were used in our study of global gauge anomalies for $SU(2)$ and other 
simple gauge groups in higher dimensions, 
in terms of Stiefel manifolds and the James numbers of the
Stiefel manifolds. For details, see refs.(7-10).

In relating to our consideration with ${(SU(2))}^3$ and $SU(8)$, 
we are interested in the case of $N=8$ from our theorem 1. 
This case $N=8$ of the theorem 1 corresponds to the fibration with base space 
$SU(2)$ and bundle space $SU(8)$, and we have  

\begin{equation}
\Pi_{5}(SU(2))\stackrel{i_*}{\longrightarrow}
\Pi_{5}(SU(8))\stackrel{{P_8}_*}{\longrightarrow}\Pi_{5}(V_6(C^8))
\stackrel{\Delta_*}{\longrightarrow}\Pi_{4}(SU(2))
{\longrightarrow}\Pi_{4}(SU(N)) = 0 ,
\end{equation}

or

\begin{equation}
0{\longrightarrow} Z\stackrel{{P_8}_*}{\longrightarrow}\Pi_{5}(V_6(C^8))
\stackrel{\Delta_*}{\longrightarrow} Z_2
{\longrightarrow} 0 ,
\end{equation}

with 

\begin{equation}
{P_8}_*(x) = 2y ,
\end{equation}
up to a torsion element of $\Pi_{5}(SU(8)/SU(2))$ and $y$ denoting the
generator of its $Z$. 

Next, in the fibration with base space ${(SU(2))}^3$ and bundle space $SU(8)$, 
with Eq.(26) for our purpose, we have
  
\begin{equation}
\Pi_{5}({(SU(2))}^3)\stackrel{{i^{\prime}}_*}{\rightarrow}\Pi_{5}(SU(8))
\stackrel{{P^{\prime}}_*}{\rightarrow}\Pi_{5}(SU(8)/{(SU(2))}^3) 
\stackrel{{\Delta^{\prime}}_*}{\rightarrow}\Pi_{4}({(SU(2))}^3)
\stackrel{}{\rightarrow}\Pi_{4}(SU(8)) = 0 .
\end{equation}
or
\begin{equation}
Z_2\oplus Z_2\oplus Z_2\stackrel{{i^{\prime}}_*}{\rightarrow} Z
\stackrel{{P^{\prime}}_*}{\rightarrow}\Pi_{5}(SU(8)/{(SU(2))}^3)
\stackrel{{\Delta^{\prime}}_*}{\rightarrow} Z_2\oplus Z_2\oplus Z_2
\stackrel{}{\rightarrow} 0 .
\end{equation}

Similarly as we have seen for Eq.(29) etc., since $Z_2\oplus Z_2\oplus Z_2$ is a
finite torsion group, the above particular ${i^{\prime}}_*$ has no non-trivial 
homomorphism, so that ${i}_* = 0$, and we can write 

\begin{equation}
0\stackrel{{i^{\prime}}_*}{\rightarrow} Z
\stackrel{{P^{\prime}}_*}{\rightarrow}\Pi_{5}(SU(8)/{(SU(2))}^3)
\stackrel{{\Delta^{\prime}}_*}{\rightarrow} Z_2\oplus Z_2\oplus Z_2
\stackrel{}{\rightarrow} 0 .
\end{equation}

We are now ready to consider homomorphism between the two homotopy
sequences Eq.(36) and Eq.(39) or Eq.(37) and Eq.(41) corresponding to 
the two fibrations,
and we obtain the commutative diagram (which hereafter will be called as
CD(I)) as follows.

$$
\CD
0 @>>> Z @>{P_8}_*>> \Pi_5(V_6(C^8)) @>>> Z_2 @>>> 0 \\ 
@. @V i VV @V q_* VV @V j_* VV\\
0 @>>> Z @>{P^{\prime}}_*>> \Pi_{5}(SU(8)/{(SU(2))}^3) @>{{\Delta}^{\prime}}_*>> Z_2\oplus Z_2\oplus Z_2 @>>> 0 
\endCD
$$

The $i$ and $j_*$ in the above commutative diagram are both a monomorphism,
by using Five Lemma$^{19}$ or Snake lemma (see also Appendix), 
$q_*$ is also a monomorphism.
A simple way to see a $j_*$ for our purpose is to note
that the top row is exact for any subgroup $SU(2)$ of $SU(8)$.
We can choose the $SU(2)$ in the top row as identical as one of the
$SU(2)$ subgroups from the 2nd row, then $j$ is an isomorphism into
a $SU(2)$ subgroup in the 2nd row or $j_*$ is an isomorphism from $Z_2$
into one of the $Z_2$'s in the 2nd row. 
 
As an abelian group, it can be written that

\begin{equation}
\Pi_{5}(SU(8)/{(SU(2))}^3) = A\oplus T^{\prime} ,
\end{equation}

where $A$ is a free subgroup and $T^{\prime}$ is the torsion 
in a canonical form (see Appendix), and up to isomorphism

\begin{equation}
A = Z^r ,
\end{equation}

for some integer r as the rank of the abelian group.

In order to determine r, we can use the tensor product$^{19}$ of the 
above commutative diagram by the additve group Q of rational numbers, 
then we have a commutative diagram (which will be called as CD(II)) as
 
$$
\CD
0 @>>> Q @>{P_8}_*>> Q @>>> 0 @>>> 0 \\
@. @V i VV @V q_* VV @V j_* VV \\
0 @>>> Q @>{P^{\prime}}_*>> Q^r @>{{\Delta}^{\prime}}_*>> 0 @>
>> 0
\endCD
$$

and therefore we have $r =1$ with $A$ being isomorphic to $Z$.
Where we have used$^{19}$ $Q\otimes T = 0$ for any torsion group $T$ and
$Q\otimes Z = Z$.

Now let $a$ be a generator of $\Pi_5(V_6(C^8))$, $b$ be a generator of
$\Pi_{5}(SU(8)/{(SU(2))}^3)$, and $c$ be a generator of $\Pi_{5}(SU(8))$. 
Then we have

\begin{equation}
q_*(a) = nb + torsion, 
\end{equation}

for some integer $n$. The commutativity of the diagram and Eq.(38) from the 
theorem 1 then gives
\begin{equation}
{P^{\prime}}_*(a) = 2nb + torsion, 
\end{equation}

We like to show that $n = \pm 1$. This can be seen easily as follows.
The image $Im {P^{\prime}}_*$ is generated by ${P^{\prime}}_*(c)$,
therefore if we have $2b\in Im {P^{\prime}}_*$, we must also have $n = \pm 1$. 
Let us note that in the 2nd row of the commutative diagram (1) as a short
exact homotopy sequence, the $Z_2\oplus Z_2\oplus Z_2$ is a quotient group of
$\Pi_{5}(SU(8)/{(SU(2))}^3)$ and it is isomorphic to$^{19}$
$\Pi_{5}(SU(8)/{(SU(2))}^3)/(Im {P^{\prime}}_*)$, i.e.

\begin{equation}
\Pi_{5}(SU(8)/{(SU(2))}^3)/(Im {p^{\prime}}_*) \cong Z_2\oplus Z_2\oplus Z_2, 
\end{equation}

and for $\forall d\in Z_2\oplus Z_2\oplus Z_2$
$2d = 0$. We especailly have 
${{\Delta}^{\prime}}_*(2b) = 2{{\Delta}^{\prime}}_*(b) = 0$ in 
$Z_2\oplus Z_2\oplus Z_2$, i.e. $2b\in Ker {{\Delta}^{\prime}}_*$ and
then it implies $2b\in Im {p^{\prime}}_*$. 
Therefore, this concludes that $n = \pm 1$ and then we can write 
\begin{equation}
q_*(a) = b + torsion .
\end{equation}

Then $q_*(a)$ can be taken as a new generator of free part of
$\Pi_{5}(SU(8)/{(SU(2))}^3)$ written as 

\begin{equation}
\Pi_{5}(SU(8)/{(SU(2))}^3) = Z q_*(a) \oplus T^{\prime} .
\end{equation}
 
We also have

\begin{equation}
{P^{\prime}}_*(c) = 2q_*(a) , 
\end{equation}
so that

\begin{equation}
\Pi_{5}(SU(8)/{(SU(2))}^3)/(Im {P^{\prime}}_*) = Z_2 \oplus T^{\prime}\cong Z_2\oplus Z_2\oplus Z_2 .
\end{equation}

This gives $T^{\prime} = Z_2 \oplus Z_2$, and therefore we have  

{\bf Theorem 2}.
\begin{equation}
\Pi_{5}(SU(8)/{(SU(2))}^3) = Z\oplus Z_2\oplus Z_2 . 
\end{equation}
     
We have also proven the following theorem.

{\bf Theorem 3}. {\it
Consider the exact homotopy sequence
\begin{equation}
\Pi_{5}(SU(8))\stackrel{P_*}{\rightarrow}
\Pi_{5}(SU(8)/{(SU(2))}^3)
\stackrel{\Delta_*}{\rightarrow}\Pi_{4}({(SU(2))}^3)
\stackrel{i_*}{\rightarrow}\Pi_{4}(SU(8)) = 0 , 
\end{equation}
or given by
\begin{equation}
Z\stackrel{P_*}{\longrightarrow}
Z\oplus Z_2\oplus Z_2
\stackrel{\Delta_*}{\longrightarrow}Z_2\oplus Z_2\oplus Z_2
\stackrel{i_*}{\longrightarrow} 0 . 
\end{equation}
Let $x\in \Pi_{5}(SU(8))$ and
$y\in \Pi_{5}(SU(8)/{(SU(2))}^3)$ be generating elements of the $Z$'s. 
Then
\begin{equation}
{P_*}(x) = 2y , 
\end{equation} 
up to a possible torsion element of $\Pi_{5}(SU(8)/{(SU(2))}^3)$.
}

We can also prove the similar topological results with $SU(2)^2$ as 
a subgroup of $SU(4)$.
Corresponding to Eq.(36) and (37), we have
\begin{equation}
\Pi_{5}(SU(2))\stackrel{i_*}{\longrightarrow}
\Pi_{5}(SU(4))\stackrel{{P_4}_*}{\longrightarrow}\Pi_{5}(V_2(C^4))
\stackrel{\Delta_*}{\longrightarrow}\Pi_{4}(SU(2))
{\longrightarrow}\Pi_{4}(SU(N)) = 0 ,
\end{equation}

or

\begin{equation}
0{\longrightarrow} Z\stackrel{{P_4}_*}{\longrightarrow}\Pi_{5}(V_2(C^4))
\stackrel{\Delta_*}{\longrightarrow} Z_2
{\longrightarrow} 0 ,
\end{equation}

with

\begin{equation}
{P_4}_*(x) = 2x ,
\end{equation}
up to a torsion element of $\Pi_{5}(SU(4)/SU(2))$ .
Corresponding to Eq.(39) or Eq.(41), we have
\begin{equation}
\Pi_{5}({(SU(2))}^2)\stackrel{{i^{\prime}}_*}{\rightarrow}\Pi_{5}(SU(4))
\stackrel{{P^{\prime}}_*}{\rightarrow}\Pi_{5}(SU(4)/{(SU(2))}^2)
\stackrel{{\Delta^{\prime}}_*}{\rightarrow}\Pi_{4}({(SU(2))}^2)
\stackrel{}{\rightarrow}\Pi_{4}(SU(4)) = 0 , 
\end{equation}
 
or

\begin{equation}
0\stackrel{{i^{\prime}}_*}{\rightarrow} Z
\stackrel{{P^{\prime}}_*}{\rightarrow}\Pi_{5}(SU(4)/{(SU(2))}^2)
\stackrel{{\Delta^{\prime}}_*}{\rightarrow} Z_2\oplus Z_2
\stackrel{}{\rightarrow} 0 .
\end{equation}

Then similar to the commutative diagrams CD(I) and CD(II), we also have

$$
\CD
0 @>>> Z @>{P_4}_*>> \Pi_5(V_2(C^4)) @>>> Z_2 @>>> 0 \\
@. @V i VV @V q_* VV @V j_* VV\\
0 @>>> Z @>{P^{\prime}}_*>> \Pi_{5}(SU(4)/{(SU(2))}^2) @>{{\Delta}^{\prime}}_*>>
 Z_2\oplus Z_2 @>>> 0
\endCD
$$

and

$$
\CD
0 @>>> Q @>{P_4}_*>> Q @>>> 0 @>>> 0 \\
@. @V i VV @V q_* VV @V j_* VV \\
0 @>>> Q @>{P^{\prime}}_*>> Q^r @>{{\Delta}^{\prime}}_*>> 0 @>
>> 0
\endCD
$$

and $r=1$. With the above needed equations and commutative diagrams given, 
the proof is 
rather similar to what we presented above with $SU(2)^3$ as a subgroup 
of $SU(8)$, the Five lemma or Snake lemma (see Appendix) will also be useful.
We will omit the further details and have the results as the
following theorems.

{\bf Theorem 4}.
\begin{equation}
\Pi_{5}(SU(4)/{(SU(2))}^2) = Z\oplus Z_2 .
\end{equation}

{\bf Theorem 5}. {\it
Consider the exact homotopy sequence
\begin{equation}
\Pi_{5}(SU(4))\stackrel{P_*}{\rightarrow}
\Pi_{5}(SU(4)/{(SU(2))}^2)
\stackrel{\Delta_*}{\rightarrow}\Pi_{4}({(SU(2))}^2)
\stackrel{i_*}{\rightarrow}\Pi_{4}(SU(4)) = 0 ,
\end{equation}
or given by
\begin{equation}
Z\stackrel{P_*}{\longrightarrow}
Z\oplus Z_2
\stackrel{\Delta_*}{\longrightarrow}Z_2\oplus Z_2
\stackrel{i_*}{\longrightarrow} 0 .
\end{equation}
Let $x\in \Pi_{5}(SU(4))$ and
$y\in \Pi_{5}(SU(4)/{(SU(2))}^2)$ be generating elements of the $Z$'s.
Then
\begin{equation}
{P_*}(x) = 2y ,
\end{equation}
up to a possible torsion element of $\Pi_{5}(SU(4)/{(SU(2))}^2)$.
}

We have now completed our proof for the topological results needed or 
used as an assumption in the section II to show the relevant $Z_2$ global
anomalies. 
Therefore, as we remarked, the relevant $Z_2$ global gauge anomalies
for the gauge groups $SU(2)^3$ in IR $\omega=(2,2,2)$ and   
$SU(2)^2$ in IR $\omega=(2,2)$ have completely proved. 
We also notice that we have obtained some additional
or more general results such as the theorem 1.  

\section{\bf Gauge Anomalies for Some Semisimple Gauge Groups in Higher Dimensions}
In this section we will discuss about the possibilities of global
 gauge anomalies for some semisimple gauge groups including also 
$SU(2)\times SU(2)$ and $SU(2)^3$ in some higher dimensions.
We will focus our discussions in $D=6, 10$ dimensions and other general $D=4k+2$ dimensions with strong anomaly-free conditions.

As we emphasized that the consideration of global gauge anomalies are meaningful
 only with the absence of local (perturbative) gauge anomalies. 
In $D=4$ dimensions, as we know this is automatically true for $SU(2)$ or its
products. 
However, local anomaly-free condition may not be true automatically 
in the higher dimensions under consideration. 

Let us now first consider the case in $D=6$ and $D=10$ dimensions.
In these dimensions, there may be generally local-anomaly involved for an IR, 
even for Lie algegras with only self-contragredient IRs. We will demonstrate
this explictly with the two semisimple gauge groups $SU(2)\times SU(2)$ and
$SU(2)^3$ as examples.
 
For Lie algebras of Lie groups $SU(2)$, $SU(3)$, $G_2$, $F_4$, $E_6$, $E_7$ and
$E_8$, as emphasized in ref.14, that they don't have a genuine $4th$-order 
Casimir invariant, and consequently for a generic element $F$ in an IR $\omega$
of the Lie algebra of the above groups, we have the trace identy given by${14}$

\begin{equation}
TrF^4 = K(\omega)(TrF^2)^2 ,
\end{equation}

where $K(\omega)$ is an overall constant depending on the representation
$\omega$, for details see ref.14. 
In particular, for $SU(2)$ in a 2-dimensional IR, we have
$K(\omega)=\frac{1}{2}$, and

\begin{equation}
TrF^4 = \frac{1}{2} (TrF^2)^2 . 
\end{equation} 

For gauge group $SU(2)\times SU(2)$ in IR $\omega=(2,2)$ in terms of dimensions, we can write,

\begin{equation}
F = F^{(1)}\otimes i_2 \oplus i_1\otimes F^{(2)},
\end{equation}

in terms of generic Lie algebric elments $F^{(1)},  F^{(2)}$  corresponding to
the two $SU(2)$'s, and the $i_1$ or $i_2$ may be regarded as the identiy
matrix corresponding to each IR space.  We then have

\begin{equation}
TrF^4 = (Tr^{(1)}{F^{(1)}}^2)^2 + (Tr^{(2)}{F^{(2)}}^2)^2) + 
6(Tr^{(1)}{F^{(1)}}^2)(Tr^{(2)}{F^{(2)}}^2) , 
\end{equation}

where the $Tr^{(1)}$ or $Tr^{(2)}$ denotes the trace in the corresponding
IR.

Therefore, the local anomaly is obviously not vanishing and the theory is
not free of local gauge anomaly for the semisimple gauge group 
$SU(2)\times SU(2)$ unless more IR(s) is to be added.

The situation is similar for the semisimple gauge group $SU(2)^3$ in the
IR $\omega=(2,2,2)$ in terms of dimensions, a similar trace identity can
also be given as

\begin{equation}
TrF^4 = 2\sum_i(Tr^{(i)}{F^{(i)}}^2)^2 + 
6\sum_{i>j}(Tr^{(i)}{F^{(i)}}^2)(Tr^{(j)}{F^{(j)}}^2) , 
\end{equation}
 
where $i, j = 1,2,3$. The theory is not free of local-anomaly unless more
IR(s) is to be added.

Similarly in $D=10$ dimensions, following ref.14 with the results above,
 we can express the local anomaly form 
$TrF^6$ for $SU(2)\times SU(2)$ in IR $\omega=(2,2)$ 
as a sum of $(Tr^{(i)}{F^{(i)}}^2)^3$,
$(Tr^{(i)}{F^{(i)}}^2)(Tr^{(j)}{F^{(j)}}^2)^2$ ($i\neq j$), 
with integer coefficients.
For $SU(2)^3$ IR $\omega=(2,2,2)$, there will be another term of
$(Tr^{(1)}{F^{(1)}}^2)(Tr^{(2)}{F^{(2)}}^2)(Tr^{(3)}{F^{(3)}}^2)$ with
integer coefficient. For more generic gauge groups, there may be terms of
higher genuine Casimir invariants.
 
Actually, there may be local-anomaly for all the simple or semisimple 
gauge groups quite generally in a generic IR in $4k+2$ dimensions. 
However, despite of this fact, for global gauge anomalies, we can have
a very generic result for groups in self-contragredient representations.
These especially include the groups with only self-contragredient 
IRs, $SP(2N)$ $(N=rank$ with $SU(2)\cong SP(2)$, $G_2$, $F_4$, $SO(2N+1)$,
$SO(4N)$, $G_2$, $F_4$, $E_7$, and $E_8$ or a product of them. 
Although the relevant homotopy groups $\Pi_{2n}(H)$ may be non-trivial,
with more than one IRs included such that the strong anomaly condition
$TrF^{n+1}=0$ is satisfied, we will conclude that 
there should be no global gauge anomaly for
gauge groups with simple ideas only listed above in arbitrary $4k+2$ dimensions.
For simple gauge groups, see details in refs.7-8, the arguments and the 
the result there should still hold even if the gauge groups are semisimple
with more than one simple ideals,
as our arguments there  assume no dependency that the gauge group is simple
although our discussions there were aimed to study with simple gauge groups.    
In fact, as emphasized in our discussions of $SU(2)$ global gauge anomalies, 
the arguments there provide a complimentary method as we used with James number
of Stiefel manifolds in other dimensions. For the semisimple gauge groups,
this is similar as the method we used in $D=4$ dimensions does not seem to 
easily apply in generic $D=4k+2$ dimensions. 
We will simply summarize it as the following result: 

{\it {\bf Proposition 1}: When the strong anomaly-free condition $TrF^{n+1}=0$
 is satisfied,
any semisimple gauge groups in a self-contragredient representation 
(which may be reducible in general) is free of global gauge anomaly in
arbitrary $4k+2$ dimensions.}

Following refs.7-8, we can also see that with the strong anomaly-free
condition for local (perturbative) gauge anomalies, any global gauge anomaly 
in arbitrary $D=2n$ dimensions is at most of a $Z_2$ or $Z_2$'s type.

We will conclude our discussions with some relevant remarks.
  
{\bf Remark (4)}: Our discussions with trace identities above are examples
involving Casimir invariants for semisimple Lie algebras. 
Obviously, as we have seen that, it in general may have mixed or crossing
invariants as product of lower invariants from different simple Lie algebras.
This is something new comparing to a simple Lie algebra 
involving Casimir invariants for semisimple Lie agebras. Casimir invariants
were extensively studied in ref.14 related to generalized Dynkin indices.

{\bf Remark (5)}: We like to remark that self-contragredient 
representations of Lie algebras were also extensively used in ref.22 for 
other stuidies, especially for unification Yang-Mills groups with CP as a
gauge symmetry. 

{\bf Remark (6)}: Our results were utilized$^1$ for the discussions of
total generation numbers of fermions (for generation) and mirror fermions 
(for antigeneration) in $SO(10)$ and
Supersymmetric $SO(10)$ unification theories, and compactification
of superstring theories on Calabi-Yau manifolds$^{20}$. 
In such superstring compactification with level-one gauge groups, 
the total number of generations obtained is even 
(see ref.20 and ref.1). 
The $SO(10)$ theories are important
models themselves for unification of fundamental forces. It is also known that
they could provide a frame in grand unified string theories (GUST)
for lifting to extent the huge degeneracy in the moduli space of parameters for
the space of classical supersymmetric string vacua, especially with gauge groups
 of level greater than one$^{21}$ to obtain odd number of total generations and
antigenerations.
There were constructions for $SO(10)$ grand unification in
four-dimensional heterotic string theory from both left-handed and right-handed
families$^{21}$. It was expected that our results$^1$ may be useful or 
providing better hints related to the relevant theories.

{\it Acknowledgement:} We like to thank B. Zhang for valuable and helpful
discussions. 

\section{\bf Appendix: Relevant Useful Theorems}

We will give here some relevant and useful theorems etc. for the convenience of
our discussions.

{\bf Five Lemma}

Given a commutative diagram of additive abelian groups and homomorphisms 

$$
\CD
G_0 @>>> G_1 @>>> G_2 @>>> G_3 @>>> G_4 \\
@V \gamma_0 VV @V \gamma_1 VV @V \gamma_2 VV @V \gamma_3 VV @V \gamma_4 VV\\
H_0 @>>> H_1 @>>> H_2 @>>> H_3 @>>> H_4 
\endCD
$$

in which each row each exact, the following holds:
(i) If $\gamma_0$ is surjective (epimorphism), and $\gamma_1$ and $\gamma_3$ are injective (monomorphism), then $\gamma_2$ is injective;
(ii) If $\gamma_4$ is injective, and $\gamma_1$ and $\gamma_3$ are surjective,
then $\gamma_2$ is surjective;

Note that when both (i) and (ii) are satisfied, the lemma also have a stronger
form.

{\bf Steenrod five lemma}

Given the above commutative diagram of additive abelian groups and homorphisms,
if $\gamma_0$,$\gamma_1$,$\gamma_3$ and $\gamma_4$ are isomorphism, then
$\gamma_0$ is isomorphism. Note that is actually a stronger case of the
lemma, and in this case both (i) and (ii) are satisfied simultaneously.
 
In the lemma, if $G_0$,$H_0$,$G_4$ and $H_4$ are trivial, and then
$\gamma_0$,$\gamma_4$ are also trivial, the lemma is in the following weaker
form:

Given a commutative diagram of additive abelian groups and homomorphisms 
$$
\CD
0 @>>> G_1 @>>> G_2 @>>> G_3 @>>> 0 \\
@.  @V \gamma_1 VV @V \gamma_2 VV @V \gamma_3 VV \\
0 @>>> H_1 @>>> H_2 @>>> H_3 @>>> 0 
\endCD
$$

in which each row each exact. 
Then $\gamma_2$ is injective (or surjective) if $\gamma_1$ and $\gamma_3$ are.
Obviously, it follows that if $\gamma_1$ and $\gamma_3$ are isomorphism, the
so is $\gamma_2$. 

{\bf Ker-coker sequence or Snake Lemma}

Given a commutative diagram of abelian groups and homomorphisms

$$
\CD
@. G_1 @>>> G_2 @>>> G_3 @>>> 0 \\
@.  @V \gamma_1 VV @V \gamma_2 VV @V \gamma_3 VV \\
0 @>>> H_1 @>>> H_2 @>>> H_3  
\endCD
$$

in which each row each exact. The there exists a homomorphism
$\Delta_*: Ker \gamma_3{\longrightarrow} Coker \gamma_1$, such that 
the following is an exact sequence: 

\begin{equation}
Ker \gamma_1{\longrightarrow} Ker \gamma_2 {\longrightarrow}Ker \gamma_3
\stackrel{\Delta_*}{\longrightarrow}
Coker \gamma_1{\longrightarrow} Coker \gamma_2 {\longrightarrow}Coker\gamma_3,
\end{equation}
where $Coker \gamma_1 =  H_1/Im\gamma_1$.
Related to our discussions, we note that if $\gamma_1$ and $\gamma_3$ are
monomorphisms, or $Ker\gamma_1=0$ and $Ker\gamma_3=0$, then $Ker\gamma_2=0$ and
$\gamma_2$ is a monomorphism.
 
{\bf Finitely Generated Abelian Groups}

{\bf Theorem} ({\it fundamental theorem of finitely generated abelian groups}).
Let $G$ be a finitely generated abelian group, and let $T$ be its torsion 
subgroup.

(i) There is a free abelian subgroup $H$ of $G$ having finite rank $\beta$ such
that $G=H\oplus T$.

(ii) There are finite cyclic groups $T_1$,...,$T_k$, where $T_i$ has order
$t_i=ord(T_i)>1$ such that $t_i$ divides $t_{i+1}$ for all i and
\begin{equation}
T = T_1\oplus ... \oplus T_k,
\end{equation}

(iii) The number $\beta$ called the {\it betti number} of $G$, and 
$T_1$,...,$T_k$ are uniquely determined by $G$.

The numbers $t_1$,...,$t_k$ are called the {\it torsion coefficients} of $G$. 

It is known that the fundamental theorem of finitely generated abelian groups
implies that any finitely generated abelian group $G$ can be written as
a finite direct sum of cyclic groups; i.e.,
\begin{equation}
G \cong (Z\oplus ... \oplus Z) \oplus Z_{t_1} \oplus ... \oplus Z_{t_k},
\end{equation}
where $t_i>1$ and $t_i$ divides $t_{i+1}$ for all i.

This representation of $G$ is also called a canonical form for $G$. 

It is also known that any finite cyclic group can be written as a
direct sum of cyclic groups whose orders are powers of primes by using
the fact that for relatively prime positive integers $p$ and $q$, we have
\begin{equation}
Z_{pq}\cong Z_p \oplus Z_q .
\end{equation}
Therefore, for any finitely generated abelian group, we can also write
\begin{equation}
G \cong (Z\oplus ... \oplus Z) \oplus Z_{a_1} \oplus ... \oplus Z_{a_s}
\end{equation}
where each $Z_{a_i}$ is a power of a prime, the numbers $a_i$ are
uniquely determined by the group $G_a$ (up to a rearrangement) and
are called the {\it invariant factors} of the group $G$.
This is another canonical form for $G$.



\vspace*{6pt}
\begin{thebibliography}{0}
\bibitem{Zhang} H. Zhang, {\it J. Group Theory Phys.} {\bf Vol.3}, {\bf No.1}, 
89 (1995) (arXiv:hep-ph/9412383), LBL-35864 (arXiv:hep-ph/9407284) 1994.
\bibitem{wit82}E. Witten, Phys. Lett. {\bf B117}, 324(1882).
 For a review of gauge anomalies, see for example, R. Rennie,
 Adv. Phys.{\bf 39}, 617(1990) and references therein.
\bibitem{wit85}E. Witten, Commun. Math. Phys.  {\bf B100}, 197. 
\bibitem{Bis86}J.-M. Bismut and D. S. Freed, Commun. Math.
 Phys. {\bf 106}, 159(1986); {\bf 107}, 103(1986); see also,
 D. S. Freed, {\it ibid.} {\bf 107}, 483(1986); S. Della
 Pietra, V. Della Pietra, and L. Alvarez-Gaume, {\it ibid.}
{\bf 109}, 691(1987); {\bf 110}, 573(1987); D. S. Freed and
 C. Vafa. {\it ibid.} {\bf 110}, 349(1987).
\bibitem{eli84}S. Elitzur and V. P. Nair, Nucl. Phys.
{\bf B243}, 205 (1984).
\bibitem{Hol86}R. Holman and T. W. Kephart, Phys. Lett. {\bf B167}, 417(1986);
 E. Kiritsis, Phys. Lett. {\bf B178}, 53(1986); {\bf 181}, 416(E)(1986).
 H. W. Braden, Univ. of North Carolina Report No. IFP-296-UNC, 
 1987, see also the review in ref.2.
\bibitem{oku88}S. Okubo, H. Zhang, Y. Tosa, and R. E.
Marshak, Phys. Rev. {\bf D37}, 1655 (1988).
\bibitem{zha88}H. Zhang, S. Okubo, and Y. Tosa, Phys. Rev.
{\bf D37}, 2946 (1988).
\bibitem{zha88}H. Zhang and S. Okubo, Phys. Rev. {\bf D38},
1800 (1988).
\bibitem{okubo89}S. Okubo and H. Zhang, in {\it Perspectives
on Particle Physics}
(ed. S. Matsuda, T. Muta, and R. Sakaki, World Scientific,
Singapore, 1989).
\bibitem{lun88}A. T. Lundell and Y. Tosa, J. Math. Phys.
{\bf 29}, 1795 (1988).
\bibitem{oku89}S. Okubo and Y. Tosa, Phys. Rev. {\bf D40},
1925 (1989).
\bibitem{James} I. M. James, Proc. London Math. Soc. {\bf 8}, 536(1958);
{\it London Mathematical Society Lecture Note Series} 
(Cambridge University Press, Cambridge, England, 1976), {\bf Vol.24}.
\bibitem{oku83}S. Okubo and J. Patera, J. Math. Phys. {\bf
24}, 2772 (1983); {\bf 25}, 219 (1984); Phys. Rev.
{\bf D31}, 2669 (1985); S. Okubo, J. Math.
Phys. {\bf 23}, 8 (1981); {\bf 26}, 2127 (1985).
\bibitem{Steenrod}N. Steenrod, {\it The Topology of Fibre
Bundles}, (Princeton Univ. Press, Princeton, NJ 1951),
G. W. Whitehead, {\it Elements of Homotopy Theory} (Springer, New York, 1978);
Nihon Sugakkai, {\it Encyclopedic Dictionary of
Mathematics}, ed. by S. Iyanaga and Y. Kawada (MIT,
Cambridge, 1977).
\bibitem{Serre}J. P. Serre, Ann. Math. {\bf 58}, 258(1953); A. Borel, Bull.
Am. Math. Soc. {\bf 61}, 397(1955).
\bibitem{gen}C. Geng, R. E. Marshak, Z. Zhao, and S. Okubo,
Phys. Rev. {\bf D36}, 1953 (1987).
\bibitem{Oku87}S. Okubo, C. Geng, R. E. Marshak, and Z. Zhao, Phys. Rev. 
{\bf D36}, 3268 (1987).
\bibitem{Mun84}J. R. Munkres, {\it Elements of Algebraic Topology} (Menlo Park,
 CA; Addison-Wesley, 1984; E. H. Spanier, {\it Algebraic Topology} (New York: 
Springer-Verlag), 1966; 
S. Maclane, {\it Categories for the Working Mathematician} 
(New York: Springer-Verlag), 1998. 
\bibitem{thfcnc}P. Candelas, G. T. Horowitz, A. Strominger, and E. Witten,
Nucl. Phys. {\bf B258}, 46(1985), see also 
B. R. Greene,
in {\it Fields, Strings, and Duality}, TASI 1996 and references therein.
\bibitem{Lewe90} D. C. Lewellen, Nucl. Phys. {\bf B337}, 61(1990); 
Z. K. Kakushadze, et. al., Phys. Rev. Lett. {\bf 77}, 2612(1996). 
\bibitem{Zhang97}H. Zhang, (arXiv:hep-ph/9509397) Int. J. Mod. Phys. A. 
vol.12, No.3, 557(1997).
\end{thebibliography}
\end{document}